\documentclass[12pt]{iopart}
\usepackage[dvipdfmx]{graphicx}
\usepackage{cite}
\usepackage{bm}

\usepackage{color}
\usepackage{iopams}  
\begin{document}
\title[Motion-induced resonance transition between hyperfine levels]{Hyperfine transition induced by atomic motion above a paraffin-coated magnetic film}

\author{Naota Sekiguchi$^{1,2}$, Hiroaki Usui$^3$, and Atsushi Hatakeyama$^2$}

\address{$^1$ Department of Electrical and Electronic Engineering, Tokyo Institute of Technology, Meguro, Tokyo 152-8550, Japan}
\address{$^2$ Department of Applied Physics, 
Tokyo University of Agriculture and Technology,
Koganei, Tokyo 184-8588, Japan}
\address{$^3$ Department of Organic and Polymer Materials Chemistry,
Tokyo University of Agriculture and Technology,
Koganei, Tokyo 184-8588, Japan}
\ead{hatakeya@cc.tuat.ac.jp}

\begin{abstract}
We measured transitions between the hyperfine levels of the electronic ground state of potassium-39 atoms (transition frequency: 460~MHz) as
the atoms moved through a periodic magneto-static field produced above the magnetic-stripe domains of a magnetic film. 
The period length of the magnetic field was 3.8~$\mu$m. The atoms were incident to the field as an impinging beam with the most probable velocity of 550~m/s and experienced a peak oscillating field of 20~mT.
Unwanted spin relaxation caused by the collisions of the atoms with the film surface was suppressed by the paraffin coating on the film. 
We observed increasing hyperfine transition probabilities as the frequency of the field oscillations experienced by the atoms increased from 0 to 140~MHz for the atomic velocity of 550~m/s, by changing the incident angle of the atomic beam with respect to the stripe domains.
Numerical calculation of the time evolution of the hyperfine states revealed that the oscillating magnetic field experienced by the atoms induced the hyperfine transitions, and the main process was not a single-quantum transition but rather multi-quanta transitions.
\end{abstract}
\vspace{2pc}
\submitto{\jpb}
\maketitle
\section{Introduction}
The interaction of atomic spin with solid surfaces is one of the central issues in the physics of atom-surface interactions. For example, spin-polarized metastable helium (He*) atoms are used to probe the magnetism of the top-most layer of 
solids~\cite{One84,Ham92,Get95,Sal96,Kur10}. The anti-spin-relaxation coatings of containers for spin-polarized gas, such as optically pumped alkali-metal vapor and noble gases, are essential when conducting experiments requiring long spin relaxation times~\cite{Rob58,Bou66,Zen83,Hel95,Bal10}. 
Magnetic fields produced above the surface by electric current or permanent magnets can control the motion of paramagnetic atoms. Two techniques often referred to are atom mirrors for reflection~\cite{Coo82,Opa92,Roa95} and atom chips for trapping~\cite{Fol00,For07,Kei16}. 

Additionally, magnetic fields produced above the surface can also induce resonance transition of the spin state, as moving atoms experience temporally oscillating magnetic fields even when the field is static. If the velocity of atoms and the spatial period of the field are well defined, well-controlled magnetic resonance transitions can be induced. The principle of this motion-induced resonance transition~\cite{Hat05} has been demonstrated using periodic current-carrying wires~\cite{Kob09,Kob10} and 
permanent magnets~\cite{Hat16, Nag17}; the latter produces stronger magnetic fields than ordinary radio-frequency and microwave fields. Notably, magnetic transitions in the sub-terahertz regime between the hyperfine levels of positroniums, with velocities on the order of $10^7$~m/s, have been realized using a microfabricated magnetic grating with a period of the order of 100~$\mathrm{\mu m}$~\cite{Nag20}; the magnetic field produced corresponded to an energy flux of 100~MW/cm$^2$ microwaves at a frequency of 200~GHz.

In principle, a wide range of frequencies are accessible in motion-induced resonance transition by varying the atoms' speed and period length. For slow atoms, decreasing the period length is a possible approach to increase the transition frequencies.
Periodic magnetic domains in a thin magnetic material can achieve a period length on the order of micrometers or less~\cite{Ger07}.
Using the micrometer periodicity produced above the surface, it is possible to induce magnetic transitions between the hyperfine states in the microwave regime ($\sim$GHz) for alkali-metal atoms with velocities of the order of $10^3$~m/s at room temperature and above.
However, collisions of atoms with the surface are inevitable, because the magnetic field is localized in the vicinity of the surface within about the periodic length, which atoms must enter.
The collisions with the surface lead to the adsorption of atoms and/or relaxation in the spin state, thus inhibiting observation of the transition~\cite{Sek18}.
Anti-spin-relaxation coatings may circumvent this problem by preventing atoms approaching the surface from being adsorbed or spin-relaxed during collisions with the surface.

In this study, we demonstrated the magnetic hyperfine transition of potassium-39 ($^{39}$K) atoms at the transition frequency of 460~MHz using the periodic magnetic domains of the surface covered with an anti-spin-relaxation coating. Our choice of magnetic material was a magnetic garnet thin film whose magnetic domains were periodic with a 3.8~$\mu$m-period-length in a stripe configuration. 
The garnet film was coated with tetracontane (C$_{40}$H$_{82}$), a type of paraffin.
The beam of potassium (K) atoms effused from an oven at 200$^{\circ}$C with the most probable velocity of 550~m/s were optically hyperfine-polarized in the upper stream from the garnet film and were then incident on the film. The hyperfine polarization of the atoms scattered from the surface was clearly dependent on the period length of the magnetic field with respect to the atoms' trajectory.
Together with the results of our numerical calculations, we concluded that the motion of impinging atoms traversing a periodic magnetic field induces transitions between the hyperfine levels.
The numerical calculations also indicated that multi-``photon'' or multi-quanta transitions occur due to a strong magnetic field near the surface.
To our knowledge, this study is the first to demonstrate atom spin manipulation using the combination of micrometer-scale magnetic domains and an anti-spin-relaxation coating.

This paper is organized as follows. The experimental setup is described in Section 2. Hyperfine-transition experimental results for the $^{39}$K beam incident on the magnetic film are presented in Section 3, together with an analysis of the magnetic field generated above the magnetic film. We discuss the observed hyperfine transitions on the basis of our calculations in Section 4. Our findings are summarized in Section 5.

\section{Experimental setup}
\begin{figure}[t]
	\centering
	\includegraphics[clip, scale=1]{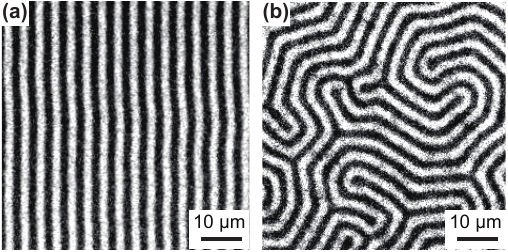}
	\caption{Magneto-optical microscope images of (a) stripe domains and (b) labyrinth domains of the garnet film.
	The dark and bright regions show opposite magnetization directions perpendicular to the surface.}
	\label{fig: domain}
\end{figure}
\begin{figure*}[t]
	\centering
	\includegraphics[clip, scale=1]{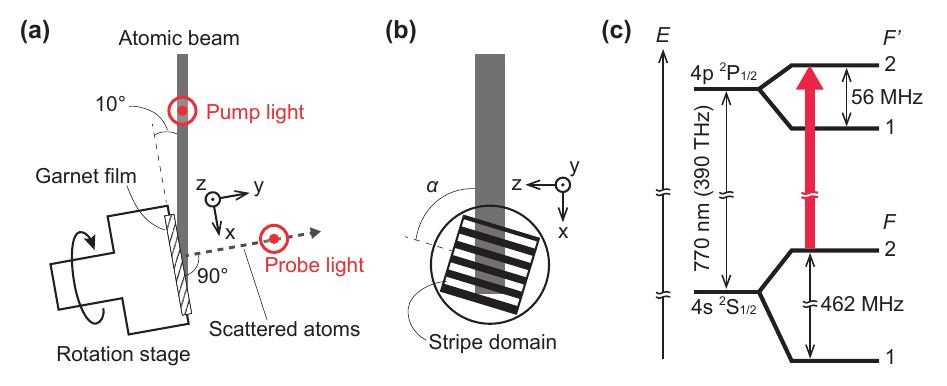}
	\caption{Experimental setup.
	(a) and (b) Schematic drawings of the apparatus from the side and front of the garnet film, respectively.
	(c) Relevant energy level diagram of $^{39}$K (not to scale).
	The red arrow represents the transition induced by the pump and probe lights.}
	\label{fig: setup}
\end{figure*}
We used a magnetic garnet film to generate a periodic magneto-static field with a period of the order of 1~$\mu$m in this work.
The garnet film was a 2-$\mu$m-thick single crystal of $\mathrm{(GdHoBi)_{3}(GaAlFe)_{5}O_{12}}$ grown on a paramagnetic gadolinium gallium garnet (GGG) substrate via liquid-phase epitaxy.
The dimensions of the garnet film were $10.5~\mathrm{mm} \times 10.5~\mathrm{mm}$.
The magnetic hysteresis curve for the garnet film had almost the same properties as reported in Ref.~\citenum{Sai16}.
The axis of easy magnetization was perpendicular to the film surface.
According to the magnetic hysteresis curve, the saturation magnetization of the garnet film was estimated to be 48~mT.
The garnet film spontaneously formed various patterns of magnetic domains with an alternating direction of the perpendicular magnetization.
We prepared two magnetic domain patterns using a strong 
magnet: a striped pattern and a labyrinth-like pattern of the magnetic domains.
Figures \ref{fig: domain}(a) and (b) show magneto-optical microscope images for the striped and labyrinth domains, respectively, used in the experiment.
The dark and bright regions indicate opposite magnetization directions perpendicular to the surface.
The period of the stripes in Fig.~\ref{fig: domain}(a) was 3.8~$\mu$m.
The paraffin film, which preserved the spin polarization of atoms as they collided with the surface, was formed on the garnet film using a vapor deposition method.
The paraffin film thickness was 0.15~$\mu$m based on atomic force microscopy measurements.

Figure~\ref{fig: setup} shows the experimental setup.
The magnetic garnet film was mounted on a rotation stage in a vacuum chamber maintained at a pressure of a few $10^{-5}$~Pa.
An atomic beam of K in natural abundance was effused from an oven heated at 200$^{\circ}$C, for which
the most probable speed of the K atoms in the beam was 550~m/s.
The atomic beam was collimated with slits (not shown) to 3~mm in one of the transverse directions of the beam (the $z$ direction in Figs.~\ref{fig: setup} (a) and (b)) and 0.7~mm in the other transverse direction (almost parallel to the $y$ direction).
$^{39}$K is the most abundant K isotope at 93\% abundance and was the main focus of the experiment.
The energies of the ground states of the $^{39}$K atom are split into hyperfine levels $F=1,2$ with 462~MHz splitting, as shown in Fig.~\ref{fig: setup}(c).

Linearly polarized pump light, tuned to the $F=2 \to F'=2$ transition of the $D1$ line, irradiated the atomic beam at normal incidence before the collisions of the atoms with the garnet film; this was done to polarize
the population of the ground hyperfine states into the $F=1$ state.
The diameter and power of the pump light were 3~mm and 50~$\mu$W, respectively.
The polarized atoms impinged obliquely on the garnet film at an incident angle of 10$^{\circ}$ with respect to the garnet surface.
For the stripe magnetic domain, we defined the angle $\alpha$ of the stripe with respect to the atomic beam direction, as shown in Fig.~\ref{fig: setup}(b). Angle $\alpha$ was variable, as the garnet film was mounted on a rotation stage.

We detected $^{39}$K atoms in the $F=2$ state scattered from the surface toward the surface normal via fluorescence induced by linearly polarized probe light using a charge-coupled device camera; the probe light was tuned to the $F=2 \to F'=2$ transition.
We note that the most probable direction of the emitted atoms is in the normal direction, as the translational motion of the colliding atoms reaches thermal equilibrium with the paraffin film; the atoms are then scattered according to the cosine law~\cite{Sek18}.
The probe light was positioned 3.5 mm from the garnet film surface.
The diameter of the probe light was about 1~mm, and the power of the probe light was 100~$\mu$W. We confirmed that the probe light power was sufficiently low so as not to affect the hyperfine polarization. A small bias magnetic field of 0.12~mT was applied to define the quantization axis ($z$ axis) along the laser propagation direction.
\section{Experiment}
\subsection{Analysis of the magnetic field generated by a garnet film}
\begin{figure*}[t]
	\centering
	\includegraphics[clip, scale=1]{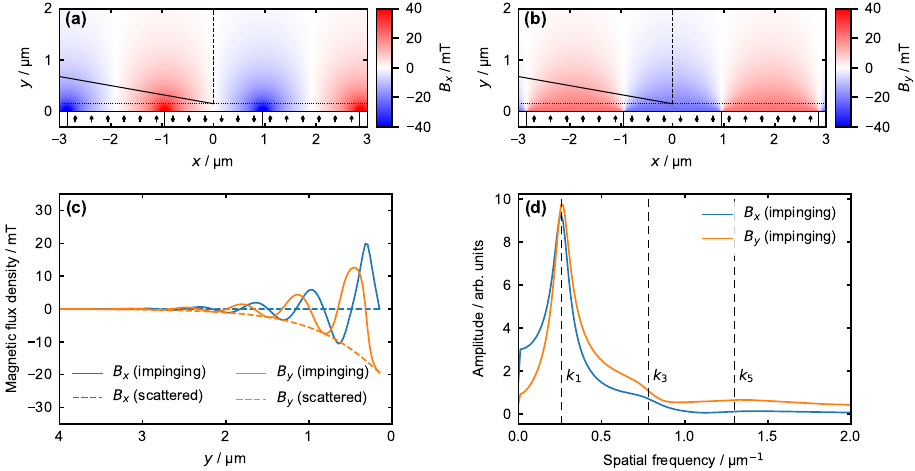}
	\caption{Magnetic field generated by the stripe domains of the garnet film.
	(a) Parallel component $B_{x}$ and (b) perpendicular component $B_{y}$ with respect to the surface. The values of the field components are indicated by the color scale bars.
	(c) Magnetic field components with respect to the atom trajectory and (d) their Fourier transform spectra for $\alpha=90^{\circ}$. The vertical dashed lines in (d) indicate first-, third- and fifth-order Fourier components. The trajectories of impinging and scattered atoms are represented by the solid and dashed lines, respectively, in (a) and (b).} 
	\label{fig: mag}
\end{figure*}
The magnetic field generated by the garnet film with stripe domains was calculated.
Figure \ref{fig: mag}(a) shows the magnetic component $B_{x}$ parallel to the surface and perpendicular to the stripes, and Fig.~\ref{fig: mag}(b) shows the normal component $B_{y}$ to the surface. The magnetic field is localized in the vicinity of the garnet-film surface and is periodic in space. In the calculation, for simplicity we assumed the stripes to be infinite in their length and coverage. The origin of the $y$ axis was located at the surface of the garnet film. The black dotted line shows the surface of the paraffin coating at $y=0.15~\mathrm{\mu m}$.
The solid lines and arrows in the garnet film ($y<0$) indicate the domain walls and magnetization directions that coincided with the calculations, respectively. The magnetization of each stripe domain was set at 48~mT (saturation magnetization).
Note that the magnetic field component along the $z$ axis is zero due to the symmetry of the stripe domains.

Figure~\ref{fig: mag}(c) shows $B_{x}$ and $B_{y}$
on the impinging trajectory, as indicated by the solid lines, and on the scattered trajectory as shown by the dashed lines in Figs.~\ref{fig: mag}(a) and (b).
The atom was assumed to collide with the paraffin film of 0.15~$\mu$m above the garnet film at $x=0$.
Oscillation of the field components occurs as the atom approaches the surface, resulting in a temporally periodic perturbation to the impinging atoms.
The Fourier transform spectra of $B_x$ and $B_y$ are shown in Fig.~\ref{fig: mag}(d);
the main (first-order) Fourier component is located at $k_1=0.26$~$\mu$m$^{-1}$, and the third- and fifth-order components are observed at about $k_3=0.78$~$\mu$m$^{-1}$ and $k_5=1.3$~$\mu$m$^{-1}$, respectively.
The frequency of the temporally oscillating magnetic field experienced by the atoms depends on $\alpha$ and the speed of the atoms. 
The main Fourier component of the oscillation was $140~\mathrm{MHz}$ for the most probable speed of 550~m/s at $\alpha=90^{\circ}$.
Rotating the stripe away from $\alpha=90^{\circ}$ resulted in a reduction in the oscillation frequency.
There was no oscillation for $\alpha=0^{\circ}$ because the impinging trajectory for $\alpha=0$ did not traverse multiple stripe domains.

The calculations 
also showed that the strength of the magnetic field reached about 20~mT on the paraffin surface, which is sufficiently strong to provide a magnetic interaction comparable to the hyperfine interaction characterized by 462~MHz-splitting. Notably, a radio-frequency wave with an energy flux of 5~MW/cm$^2$ produces an oscillating magnetic field with an amplitude of 20~mT.
\subsection{Measurement of the hyperfine transition}
We investigated changes in the populations of the hyperfine levels of $^{39}$K atoms as they scattered from the tetracontane-coated garnet 
film.
The pump light polarized most populations of incident atoms into the $F=1$ state.
We detected the fluorescence $I_{\mathrm{p}}$ from the scattered atoms using the probe light.
The fluorescence $I_{\mathrm{0}}$ from the scattered atoms in the absence of the pump light, that is, without polarization of the hyperfine-level population, was also recorded.
We then evaluated the difference $\Delta_{\mathrm{fl, s}}$ in fluorescence signals from the scattered atoms to be
\begin{equation}
\Delta_{\mathrm{fl, s}} = 
\frac{I_{\mathrm{0}} - I_{\mathrm{p}}}
{I_{\mathrm{0}}}.
\end{equation}
The fluorescence difference for the impinging atoms $\Delta_{\mathrm{fl, i}}$, was measured independently in the same manner.
The ratio $P$ between the fluorescence differences before and after scattering from the garnet film represents the ratio of the differences in the population~\cite{Sek18},
\begin{equation}
P = \frac{\Delta_{\mathrm{fl, s}}}{\Delta_{\mathrm{fl, i}}}
=\frac{
5 n_{\mathrm{1, s}} - 3n_{\mathrm{2, s}}
}{
5 n_{\mathrm{1, i}} - 3n_{\mathrm{2, i}}
},
\end{equation}
where $n_{F, \mathrm{i}}$ and $n_{F, \mathrm{s}}$ ($F=$1 or 2) are the populations of impinging and scattered atoms, respectively, at the hyperfine level specified by $F$ in the presence of the pump light. Prefactors 3 and 5 represent the statistical weights of $F=1$ and 2, respectively.
The ratio $P$ reflects the transitions between the hyperfine levels during scattering from the garnet film. $P$ is equal to 1 when the initial hyperfine polarization is preserved during scattering ($n_{1,s}/n_{2,s}=n_{1,i}/n_{2,i}$), while $P$ is equal to 0 when the hyperfine polarization is completely lost and $n_{1,s}/n_{2,s}$ becomes $3/5$.
Hereinafter, we refer to $P$ as ``surviving hyperfine polarization'' in this sense.
We measured the dependence of $P$ on the stripe angle $\alpha$.
This dependence on $\alpha$ should originate from changes in the frequency of the oscillating magnetic field experienced by the atoms.
\begin{figure}[t]
	\centering
	\includegraphics[clip,scale=1]{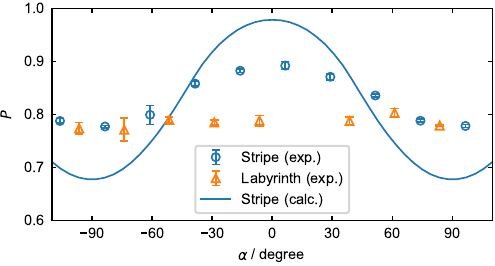}
	\caption{Surviving hyperfine polarization $P$ as a function of $\alpha$.
	The open circles and triangles show measured $P$ for the stripe and labyrinth domains, respectively.
	The solid curve shows the numerically calculated $P$ (see Sec.~\ref{sec: calc}).}
	\label{fig: survived}
\end{figure}

The measured surviving hyperfine polarizations $P$ for the stripe-domain garnet are shown by the open circles in Fig.~\ref{fig: survived} as a function of $\alpha$.
The error bars indicate standard deviations of the mean of five measurements at each $\alpha$.
We found that the surviving hyperfine polarization clearly depends on $\alpha$ with minimums at $\alpha=\pm90^{\circ}$, where the hyperfine transition occurs with the highest probability.
This dependence agrees qualitatively with the fact that the effective period length for the impinging atoms became shorter, and the frequency experienced by the atoms approached the hyperfine frequency as the stripe rotated to $\alpha=\pm90^{\circ}$.

We also performed the same measurement for the labyrinth-domain garnet. As shown in Fig.~\ref{fig: survived} with the open triangles, 
in contrast to the stripe domain case, the surviving hyperfine polarization was constant over the range of $-90^{\circ} \lesssim \alpha \lesssim 90^{\circ}$ for the labyrinth domains.
No dependence on $\alpha$ in the case of the labyrinth domain indicated that the rotation of the stripe domains was essential for observing the $\alpha$ dependence of $P$, as opposed to the rotation of the mount or GGG substrate.
Although the constant polarization for the labyrinth domain pattern should theoretically be equal to the polarization averaged over $\alpha$ for the stripe domain pattern, we consider that it was not equal in our experiment due to some experimental conditions that changed during the process of changing the shape of the magnetic domains, including opening and re-evacuating the vacuum chamber, and dismounting and re-mounting the garnet film.
\section{Discussion}
\label{sec: calc}
\begin{figure}[t]
	\centering
	\includegraphics[clip, scale=1]{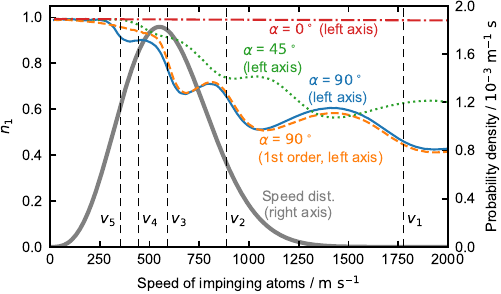}
	\caption{Speed distribution of the impinging atoms and dependence of the calculated population in the $F=1$ state $n_{1}$ after scattering with respect to the speed of the impinging atoms. The vertical dashed lines indicate velocities for $n$-quanta transitions ($n=1, 2, 3, 4$, and 5).}
	\label{fig: speed}
\end{figure}
In an attempt to better understand our experimental results and the underlying mechanisms, we calculated the spin state of atoms that impinged on the oscillating field produced by the garnet film.
We computed the time evolution of the density matrix $\rho$ that follows from Liouville's equation:
\begin{equation}
\frac{\partial \rho}{\partial t}=\frac{1}{i \hbar}\left[ H, \rho \right],
\label{eq: liouville}
\end{equation}
using eight Zeeman sublevels of the hyperfine states of $^{39}$K as bases.
Here, the Hamiltonian $H$ for the atoms is given by
\begin{equation}
H(t) = \frac{A}{2\hbar^{2}}\bm{S}\cdot\bm{I} - \bm{\mu}\cdot \bm{B} (t).
\label{eq: ht}
\end{equation}
The first term represents the hyperfine interaction with the dipole coupling coefficient $A$ for the $S_{1/2}$ state of $^{39}$K, the reduced Planck's constant $\hbar$, the electron spin $\bm{S}$, and the nuclear spin $\bm{I}$.
The second term represents the interaction of the atomic magnetic moment $\bm{\mu}$ with the magnetic field $\bm{B}(t)$ experienced by the 
atoms.
The magnetic moment was calculated as follows:
\begin{equation}
\bm{\mu}=-g_{S}\mu_{B}\bm{S}+\frac{\mu_{I}}{I}\bm{I},
\label{eq: mu}
\end{equation}
where $g_{S}$ is the Land\'{e} g-factor, $\mu_{B}$ is the Bohr magneton, $\mu_{I}$ is the nuclear magneton, and $I$ is the nuclear spin quantum number; $I=3/2$ for $^{39}$K.
We evaluated the population $n_{1}$ at the $F=1$ level.
$n_1=1$ was set as the initial value, that is, all of the population is in the $F=1$ level before the collision with the garnet film,
and we calculated $n_1$ 5~$\mu$m above the paraffin surface for atoms scattered in the direction normal to the surface. Calculations were performed without dwell times for the atoms on the paraffin surface; we confirmed that finite dwell times did not affect our calculation results.

The magnetic field $\bm{B}(t)$ for a given stripe angle $\alpha$ depends on the speed and trajectory of the atoms.
To take into account the width of the atomic beam, which was much larger than the stripe period, we considered various trajectories with different collision points on the paraffin coating.
We then took an average of $n_{1}$ over the collision points within one stripe period.
The distribution $f_{v}$ of the speed $v$ of atoms impinging on and scattering from the garnet film in the direction of the surface normal is given by
\begin{equation}
f_{v}(v)dv = \frac{2 v^{3}}{v_{D}^{4}}\exp
\left(
-\frac{v^{2}}{v_{D}^{2}}
\right)
dv.
\end{equation}
Here, $v_{D}=\sqrt{2k_{B}T/m}$, with Boltzmann's constant $k_{B}$, temperature $T$ of the atomic gas, and atomic mass $m$ of $^{39}$K~\cite{Ram56}.
The temperature of the impinging atoms was considered to be the same as the oven temperature of 473~K, while the temperature of the scattered atoms was estimated as 300~K due to thermalization with the paraffin coating at room temperature~\cite{Sek18}.
The most probable speed is given by $\sqrt{3/2}v_D$, being 550~m/s and 440~m/s for the impinging and scattered atoms, respectively. The velocity distribution of the impinging atoms is plotted in Fig.~\ref{fig: speed}.

\begin{figure}[t]
	\centering
	\includegraphics[clip, scale=1]{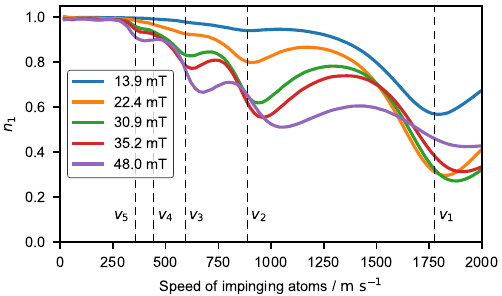}
	\caption{Surface magnetization dependence of the calculated population in the $F=1$ state $n_{1}$ after scattering with respect to the speed of the impinging atoms. The vertical dashed lines indicate velocities for $n$-quanta transitions ($n=1, 2, 3, 4$, and 5).}
	\label{Msdep}
\end{figure}

The calculated value for $n_{1}$ is shown Fig.~\ref{fig: speed} as a function of the injection speed of the atoms. $n_{1}$ was averaged over the speed distribution of the scattered atoms.
The vertical dashed lines indicate velocities $v_n=462$~MHz$/(nk_1)$ ($n=$1, 2, 3, 4, or 5), at which magnetic transitions similar to transitions induced by $n$-photons with a frequency of $v_nk_1$ can occur. We call these transitions with $n\ge 2$ multi-quanta transitions in this paper, while the transition with $n=1$ is called single-quantum transition.
We attributed the deepest dip in $n_{1}$ for $\alpha=90^{\circ}$ (blue curve) around 1$\,$900~m/s to a single-quantum magnetic transition between the hyperfine levels.
However, this single-quantum transition should not contribute significantly to the experimentally observed hyperfine transitions because the beam contained few atoms with velocities of about 1$\,$900~m/s.
Other dips in $n_{1}$ for $\alpha=90^{\circ}$ around 1$\,$100~m/s and 650~m/s were major contributors to the observed transition due to the relatively large number of atoms in the impinging beam. These dips are considered to be resonances with two and three quanta, respectively.
Multi-quanta transitions, not between hyperfine levels but between Zeeman sublevels, have been investigated theoretically and experimentally~\cite{Shi65, Ste72a, Coh98, Sun22}.
We conducted calculations with varying saturating magnetization of the garnet film and found that the dips around two- and three-quanta transitions became more prominent as the magnetization of the garnet exceeded 10 mT, which resulted in a magnetic field of several mT on the paraffin surface, as shown in Fig.~\ref{Msdep}.
The strong oscillating field on the order of 10 mT at the paraffin surface, realized in this experiment with a magnetization of 48~mT of the garnet film, induced multi-quanta magnetic transitions from the prepared $F=1$ state. 

To check the degree to which the higher-order Fourier components of the oscillating field contributed to these dips, we calculated $n_1$ for a purely sinusoidal magnetic field at the first-order spatial frequency, which was generated by the first-order Fourier component of the surface magnetization distribution.
We confirmed that the main dips at 1$\,$900, 1$\,$100, and 650~m/s were reproduced with this sinusoidal field as shown in Fig.~\ref{fig: speed} (orange curve). Notably, the minor structure from 300 to 600~m/s was not reproduced. We consider that this structure corresponds to a single-quantum resonance with the third-order component ($k_3$)  and fifth-order component ($k_5$) of the magnetic field, which are shown in Fig.~\ref{fig: mag}(d). We therefore concluded that higher-order components of the magnetic field produced by the garnet film induced the hyperfine transitions, but they had a relatively minor contribution.

To better understand the dependence of $n_1$ on the field frequency, we calculated $n_1$ for various stripe angles. The curve for $\alpha=45^{\circ}$ (green curve) in Fig.~\ref{fig: speed} shows that the dips reasonably shift to higher velocities by a factor of $\sqrt{2}$ compared to $\alpha=90^{\circ}$, and they generally become small as well. Finally, the dependence for $\alpha=0^{\circ}$ (red curve) shows 
no dips.
The results of these calculations further confirm that the oscillating magnetic field experienced by the atoms induces hyperfine transitions. 

On the basis of the incident speed of the atoms and the stripe angle dependence, as discussed above, we calculated $P$ for the stripe domain, where the initial population of impinging atoms was assumed from the measured value of $\Delta_{\mathrm{fl, i}}=0.87$.
We took a weighted average of the calculated $n_{1}$ with the impinging speed distribution and evaluated $P$ based on the averaged $n_{1}$.
The calculated stripe-angle dependence shown in Fig.~\ref{fig: survived} agreed reasonably well with the measured dependence. The calculation may be improved by considering multiple factors, including the details of the garnet film domains, the imperfection of the paraffin coating, and the temporal change in magnetic field experienced by atoms diffusing on the surface.
We therefore concluded that the observed decrease in $P$, or the observed increase in the hyperfine transition, at $\alpha =\pm 90^{\circ}$ was caused by the oscillating magnetic field experienced by the impinging atoms. The main processes were multi-quanta magnetic transitions.
\section{Conclusions}
We investigated the hyperfine transitions of $^{39}$K atoms that impinged on a periodically magnetized garnet film coated with paraffin, which prevented unwanted adsorption and spin relaxations of the atoms on the surface.
The magnetic domain of the garnet film had a striped configuration that generated a periodic magneto-static field above the paraffin coating.
The impinging atoms experienced a temporally oscillating magnetic field, the strength of which reached the order of 10 mT.
We observed that the hyperfine polarization of $^{39}$K atoms scattered from the garnet film depended on the incident angle with respect to the stripe direction.
Taken together with the results from the numerical calculations, the hyperfine transitions occurred when the impinging atoms traversed the stripe domain and experienced the oscillating magnetic field.
Furthermore, the numerical calculation confirmed multi-quanta magnetic resonances between the ground hyperfine levels.
This study thus demonstrates that micrometer-scale magnetic domains of magnetic materials can be used to manipulate the spin states of impinging atoms through motion-induced resonance transitions when the surface is covered with an anti-relaxation coating to suppress unwanted spin relaxation at the surface.

\section*{Acknowledgments}
We thank N. Matsuzaka for his technical support. This work was supported by JSPS KAKENHI Grant Numbers JP17J03089, JP17H02933 and JP23H01845.

\providecommand{\newblock}{}

\end{document}